\documentclass[journal=aesccq,manuscript=letter,layout=twocolumn]{achemso}
\usepackage{chemformula} 
\usepackage[T1]{fontenc} 

\author{Gerard Meijer}
\affiliation[FHI Berlin]{Fritz Haber Institute of the Max Planck Society, Faradayweg 4-6, 14195 Berlin, Germany}
\email{meijer@fhi-berlin.mpg.de}
\author{Gert von Helden}
\affiliation[FHI Berlin]{Fritz Haber Institute of the Max Planck Society, Faradayweg 4-6, 14195 Berlin, Germany}

\title[M-MS-S-SC-C type stars]
  {Metal monoxide abundances as a function of the C/O ratio}

\abbreviations{IR,NMR,UV}

\begin{document}

\begin{abstract}
  The diatomic metal monoxides whose optical spectra define the classification of stars on the asymptotic giant branch (AGB), that is, TiO, YO, ZrO, and LaO, have the unusual property that their ionization energy is below their dissociation limit. The cations of these metal monoxides can be efficiently produced via associative ionization of their constituent ground state atoms and are long-lived. We present a simple model that can explain the observed relative abundance of these metal oxides as a function of the C/O ratio.
\end{abstract}

\section{Introduction}
In a possibly seminal but still largely overlooked paper, \citeauthor{Schofield2006} pointed out that there is a special series of 18 metal elements that can have long-lived metal oxide cations \cite{Schofield2006}. These cations have the unusual property that their ground state is energetically below the dissociation limit of the neutral molecule. This makes these cations stable against dissociative recombination with electrons, the common destruction process for cations. This same property also enables efficient production of these cations from their constituent ground state atoms via associative ionization, also called chemi-ionization, without additional collision energy. These cations can be neutralized by charge exchange with atoms that have ionization energies below that of the metal monoxides. Interestingly, the neutral metal monoxides listed in the catalog of stars that provide standards of spectral type while the stars evolve along a sequence from type M via MS through S and SC to the carbon stars, namely TiO, YO, ZrO, and LaO, all belong to this select, special series \cite{Keenan1980}. The change in the relative abundance of these metal oxides is due in part to the increase in the abundance of heavy elements while the stars evolve \cite{Abia1998}. The relative abundance of these metal monoxides is also thought to correlate with the simultaneously increasing C/O ratio and changes remarkably in a narrow range of C/O ratios just below 1.0: TiO dominates when C/O $<$ 0.95, TiO and ZrO are comparable when C/O = 0.96 and then also YO is strong, ZrO becomes dominant for values of C/O of 0.97-0.98, while LaO is most prominent for values of C/O even closer to one \cite{Keenan1980}.

\section{Results and discussion}

When associative ionization of a metal atom, M, and an oxygen atom, O, is energetically possible, the 
\begin{equation}
    \text{M}+\text{O} \rightarrow \text{MO}^+ +\text{e}^- 
\end{equation}
two-body reaction with a rate constant $k_{ai}$ will dominate the production of the metal monoxide cation, MO$^+$. The strongly bound MO$^+$ cation can be destructed via abstraction reactions with neutral atoms, provided the resulting neutral diatomic oxide is even more strongly bound. Given that CO has a binding energy of 11.1 eV, oxygen abstraction with carbon atoms via 
\begin{equation}
    \text{MO}^+ + \text{C} \rightarrow \text{M}^+ + \text{CO} 
\end{equation}
with a rate constant $k_{oa}$ will dominate the depletion of MO$^+$. In equilibrium, we can write: 
 \begin{equation}
 [\text{MO}^+]=\frac{k_{ai} [\text{M}]_f [\text{O}]_f}{k_{oa} [\text{C}]_f}=\frac{k_{ai}[\text{O}]_f}{k_{oa}[\text{C}]_f + k_{ai}[\text{O}]_f}[\text{M}]_t
 \end{equation}
 where the amount of the species is given in square brackets and where [M]$_t$$\equiv$[M]$_f$+[MO$^+$] has been used. The subscript $f$ is used to indicate that these are the free M, O and C atoms. In our model, we assume that all C and O atoms are bound in CO but that the surplus O and C atoms are free. In an environment with a C/O ratio smaller than one, [O]$_f$ is thus taken to be equal to [O] - [C] and [C]$_f$=0; when the C/O ratio is larger than one, [C]$_f$ is taken to be equal to [C] - [O] and [O]$_f$=0. Using only this, Eq. (3) would lead to a Heaviside step function for [MO$^+$]/[M]$_t$, i.e. its value would be 1 for C/O$<$1 and 0 for C/O$>$1 for all M. 

 It appears realistic to assume that there will always be some free C and O atoms around, because in kinetic equilibrium a fraction $\alpha$ of the CO molecules will be (photo)-dissociated. When C/O$\leq$1, we can take [C]$_f$=$\alpha$[C] while [O]$_f$ will not get below $\alpha$[C] either and we find: 
 \begin{equation}
 \frac{[\text{MO}^+]}{[\text{M}]_t}=\frac{k_{ai}([\text{O}]-[\text{C}]+\alpha[\text{C}])}{k_{oa}\alpha [\text{C}] + k_{ai}([\text{O}]-[\text{C}]+\alpha[\text{C}])}
 \end{equation}
 When C/O$\geq$1, we can take [O]$_f$=$\alpha$[O] while [C]$_f$ will not get below $\alpha$[O] and we find: 
 \begin{equation}
 \frac{[\text{MO}^+]}{[\text{M}]_t}=\frac{k_{ai}\alpha [\text{O}]}{k_{oa}([\text{C}] -[\text{O}]+\alpha[\text{O}]) + k_{ai}\alpha [\text{O}]}
 \end{equation}

As seen in the last column of Table~\ref{tbl:Table I}, the values of $k_{ai}$, taken from from Ref. \citenum{Schofield2006}, vary over more than two orders of magnitude for M=Ti, Y, Zr, and La. In view of the large and similar exothermicity of reaction (2) for these four metal atoms (2.5--4.3 eV; see $D_0$(MO$^+$) values in Table~\ref{tbl:Table I}), their $k_{oa}$ values are expected to be similar. We can take $k_{oa}$ as the classical Langevin rate constant for ion-neutral collisions as 2.34 x 10$^{-9}\sqrt{1.76/\mu}$ cm$^3$ s$^{-1}$, where 1.76 \AA$^3$ is the polarizibility of the C-atom and $\mu$ is the reduced mass (in amu) of the C-MO$^+$ collision complex. The parameter $\alpha$ only depends on the environment, not on M. 

\begin{table*}
  \caption{For MO molecules with M=Ti, Y, Zr, and La, the electronic ground state X(MO), dissociation energy $D_0$(MO) and ionization energy $IE$(MO) is given followed by the electronic ground state X(MO$^+$) of the cation and its $D_0$(MO$^+$) value \cite{Jain2023,Huang2013,Linton1999,Luo2015,Cao2021}. The values of $\Delta D_0$ $\equiv$ $D_0$(MO$^+$)-$D_0$(MO) = $IE$(M)-$IE$(MO) are more accurately known than either $D_0$(MO) or $D_0$(MO$^+$). All values are in eV. The rate constant $k_{ai}$ for the associative ionization reaction (Eq. (1)) is given in cm$^3$ s$^{-1}$ (from Ref. \citenum{Schofield2006}).}
  \label{tbl:Table I}
\begin{tabular}{||l||l|l|l||l|l||l||l||}
	\hline
	 MO	& X(MO) & $D_0$(MO) & $IE$(MO) 	& X(MO$^+$) & $D_0$(MO$^+$) & $\Delta D_0$ & $k_{ai}$(298 K) \\	
	 \hline
	 TiO & X$^3\Delta$ & 6.824(10) 	& 6.8198(1) & X$^2\Delta$ &  6.832(10) & 0.0084(1) & 1.7 x 10$^{-12}$ \\
      YO	& X$^2\Sigma^+$ & 7.14(18)		& 6.112958(4)  & X$^1\Sigma^+$ &  7.24(18)	& 0.1041(1) & 1.5 x 10$^{-11}$\\
	 ZrO	& X$^1\Sigma^+$ & 7.94(11)		& 6.81272(10) & X$^2\Delta$ &  7.76(11)   & -0.1788(1) & 6.0 x 10$^{-11}$ \\
	 LaO	& X$^2\Sigma^+$ & 8.23(9) & 5.2446(6)    & X$^1\Sigma^+$ &  8.57(9) & 0.3324(8) & 3.5 x 10$^{-10}$ \\
	\hline
\end{tabular}\\
\end{table*}

\begin{figure}
\includegraphics[width=8.3cm]{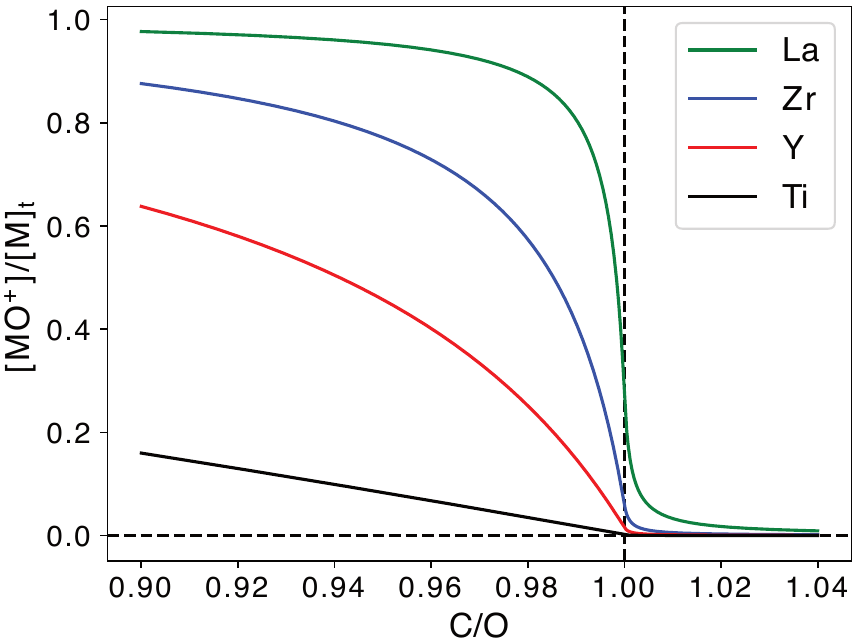}
  \caption{Plot of the relative abundance [MO$^+$]/([M]$_f$+[MO$^+$]) as a function of the C/O ratio for M=Ti, Y, Zr, La according to eq. (4) and (5) for $\alpha$=10$^{-3}$. The ratio $k_{ai}$/$k_{oa}$ is 1.70x10$^{-3}$, 1.57x10$^{-2}$, 6.29x10$^{-2}$, and 3.75x10$^{-1}$ for M=Ti, Y, Zr, and La, respectively.}
  \label{fgr:Meijer-Figure1}
\end{figure}

In Figure 1, the expressions (4) and (5) are plotted for M=Ti, Y, Zr, and La as a function of the C/O ratio, from 0.90-1.04, for a value of $\alpha$=10$^{-3}$. When C/O is close to, but just below 1.0, the relative abundance [MO$^+$]/[M]$_t$ depends approximately linearly on the product of $k_{ai}$ and (1-C/O). It also depends inversely proportional on the product of $k_{oa}$ and $\alpha$, but the latter product is basically the same for all M. The relative abundance of LaO$^+$ increases rapidly when C/O is reduced from 1.0, while the free La atoms get rapidly depleted. The relative abundance of LaO$^+$ already reaches its maximum for C/O values just below 1.0, and this goes faster the smaller the value of $\alpha$. The relative abundance of ZrO$^+$ grows slower with a decreasing C/O ratio, followed by YO$^+$ that approaches its maximum at still lower C/O values. Obviously, the relative abundance [MO$^+$]/[M]$_t$ needs to be multiplied by the actual abundance [M] of the metal to find the actual abundance [MO$^+$]. Although [Ti] is considerably larger than either [Y], [Zr], or [La] the TiO$^+$ cation will only become dominant at yet lower values of C/O as associative ionization of Ti+O is near-thermo-neutral\cite{Jain2023}, and about a factor 200 slower than for La+O. In our sun, [Ti] is about two orders of magnitude larger than [Zr]. It has been observed that in SC-type stars, the abundance of s-process elements like Zr is increased by about one order of magnitude relative to our sun \cite{Abia1998}. The value of $\alpha$ has been chosen such that when the [Ti]/[Zr] ratio of about 10 for these stars is included, the curves for Ti and Zr as shown in Figure 1 intersect around a C/O = 0.96 ratio \cite{Keenan1980}.

What is astronomically observed is the intensity of the optical absorption of the neutral MO species, which correlates with their abundance, although not necessarily in the same way for the different species. The ground-state partition functions, the oscillator strengths of the rotational branches of the electronic transitions and their isotopic pattern are likely to enhance the optical absorption of LaO and YO, and probably also of ZrO, compared to that of TiO beyond the actual abundances. The abundance of the neutral MO species will be correlated with the abundance of the MO$^+$ cations. Therefore, the intensity of the optical absorption of the neutral MO species correlates with the abundance of the MO$^+$ cations, and the behaviour as a function of the C/O ratio as described above and shown in Figure 1 should also hold for the intensity of the absorption of the neutral MO species. 

\citeauthor{Schofield2006} has reported that the values for $k_{ai}$ do not appear to be sensitive to temperature \cite{Schofield2006}. However, the $k_{ai}$ values are seen to increase with increasing energy difference between the dissociation limit and the ionization energy for the different metal monoxides. Based on this observation one could expect the rate constants for associative ionization to also increase with temperature. Such an overall increase with temperature would not influence our conclusions regarding the relative abundance of the various metal monoxides as a function of the C/O ratio.

Associative ionization with atomic oxygen is also exothermic for Sc and Ce, and both ScO and CeO have been observed in M-type stars. However, the value of $k_{ai}$ for the production of CeO$^+$ is only four times larger than for TiO$^+$ and given the lower abundance of Ce, one would not expect the CeO bands to become dominant for any C/O ratio \cite{Schofield2006}. The value of $k_{ai}$ for the production of ScO$^+$ is not known. For vanadium atoms the associative ionization reaction with oxygen atoms is endothermic by 0.693(2) eV \cite{Merriles2020}. Absorption bands of VO were inconspicuous at first in M-type stars, which was taken as an indication that the dissociation energy for VO is lower than for TiO. Although correct, this difference is now known to be only 0.28 eV \cite{Jain2023,Merriles2020}, and instead, the endothermicity of the associative ionization reaction might be of importance here. The latter would also explain the extreme temperature sensitivity of the VO absorption bands that have been observed in M-type stars \cite{Hearnshaw2014}. 

The dominance of the TiO, YO, ZrO, and LaO species has been rationalised in the chemical models thus far mainly by their large binding energies. Boron monoxide (BO) is known to have an even larger binding energy of about 8.4 eV \cite{Magoulas2014} but its well-known visible absorption bands \cite{Jenkins1932} have not been reported upon in stellar spectra, notwithstanding that Bobrovnikoff listed BO among the twelve metal oxides whose bands have been observed \cite{Bobrovnikoff1939}. As the ionization potential of BO is about 4.5 eV above its dissociation limit \cite{Magoulas2014}, associative ionization cannot play a role at all in the production of BO$^+$, which might explain the lack of data on the absorption of BO in the stellar spectra.

\section{Conclusions}

With the simple model presented here, we want to stress the importance of associative ionization reactions of a select series of 18 metal atoms with atomic oxygen \cite{Schofield2006,Oppenheimer1977}. For the metal atoms discussed here, these reactions have largely different rate constants and can thereby explain the observed intensity of the absorption of the MO species as a function of the C/O ratio for AGB stars along the M-MS-S-SC-C sequence. It can also explain that in stars that are particularly rich in ZrO, the atomic lines of Zr are entirely absent, i.e. that Zr is wholly oxidized, whereas Ti is not, an observation that was made as early as 1934 \cite{Bobrovnikoff1934}. The steep decrease of the Zr and La atom abundance for a C/O ratio close to 1.00 can also explain the observed large fluctuations in the measured values of [Zr] and [La] for stars that have a C/O ratio that is nominally 1.00, but that will differ slightly ($\approx$0.01) around this value from star to star \cite{Abia1998}. 

In many parts of the stellar and interstellar regions the chemistry is characterized by the formation and destruction of molecules. To model these processes, complex reaction networks have been set up. Including associative ionization in these reaction networks might provide a more accurate description and yield a better understanding of these environments. 

\begin{acknowledgement}
The authors acknowledge inspiring discussions with Harold Linnartz on this topic. 
\end{acknowledgement}


\providecommand{\latin}[1]{#1}
\makeatletter
\providecommand{\doi}
  {\begingroup\let\do\@makeother\dospecials
  \catcode`\{=1 \catcode`\}=2 \doi@aux}
\providecommand{\doi@aux}[1]{\endgroup\texttt{#1}}
\makeatother
\providecommand*\mcitethebibliography{\thebibliography}
\csname @ifundefined\endcsname{endmcitethebibliography}
  {\let\endmcitethebibliography\endthebibliography}{}



\end{document}